\documentclass[preprint,aps,pre,showpacs,superscriptaddress,nofootinbib]{revtex4}
\usepackage{graphicx}
\usepackage{dcolumn}
\usepackage{bm}

\def\be{\begin{equation}}
\def\ee{\end{equation}}
\def\bea{\begin{eqnarray}}
\def\eea{\end{eqnarray}}
\def\bi{\begin{itemize}}
\def\ei{\end{itemize}}

\input{tcilatex}
\begin{document}

\preprint{\today}
\title{ Scaling Relations for Collision-less Dark Matter Turbulence}
\author{Akika Nakamichi \thanks{%
akika@astron.pref.ginma.jp}}
\affiliation{Gunma Astronomical Observatory, 6860-86 Nakayama, Takayama Agatsuma, Gunma
377-0702, Japan}
\author{Masahiro Morikawa \thanks{%
hiro@phys.ocha.ac.jp}}
\affiliation{Department of Physics, Ochanomizu University, 2-1-1 Ohtuka, Bunkyo, Tokyo
112-8610, Japan}

\begin{abstract}
Many scaling relations are observed for self-gravitating systems in the
universe. We explore the consistent understanding of them from a simple
principle based on the proposal that the collision-less dark matter fluid
terns into a turbulent state, i.e. dark turbulence, after crossing the
caustic surface in the non-linear stage. The dark turbulence will not eddy
dominant reflecting the collision-less property. After deriving Kolmogorov
scaling laws from Navier-Stokes equation by the method similar to the one
for Smoluchowski coagulation equation, we apply this to several observations
such as the scale-dependent velocity dispersion, mass-luminosity ratio,
magnetic fields, and mass-angular momentum relation, power spectrum of
density fluctuations. They all point the concordant value for the constant
energy flow per mass: $0.3 cm^2/sec^3$, which may be understood as the speed
of the hierarchical coalescence process in the cosmic structure formation.
\end{abstract}

\maketitle


Cosmology: theory, Gravitation, Turbulence, Cosmology: dark matter,
Cosmology: observations, Hydrodynamics 
98.80.-k 
95.35 +d 
98.65.-r 


\section{Introduction}

Why the stars and galaxies are rotating? What is the origin of the angular
momentum? It is known that the planets in our solar system show a remarkable
scaling relation between the angular momentum $J$ and the mass $M$ for each
planet, 
\begin{equation}
J\simeq \alpha \frac{G}{c}M^{2}=\alpha \left( {\frac{\hbar }{m_{pl}^{2}}}%
\right) M^{2},\quad \alpha \simeq 10^{4}  \label{JMplanets}
\end{equation}
where the unit mass and the angular momentum are measured respectively by
the Planck mass $m_{pl}\equiv \left( c\hbar /G\right) ^{1/2}=2.18\times
10^{-5}$g\ and the reduced Planck constant $\hbar \equiv 1.05\times 10^{-27}$%
erg$\cdot $sec. Only exceptions are the Mercury and the Venus whose
rotations are locked with their orbital rotation around the Sun. Although
the Sun by itself deviates from the above scaling, the whole solar system is
also on this line\footnote{%
If the above scaling is universally holds, then in general, the slowlly
rotating star has a larger possibility to harbor a planet system since the
most of the angular momentum is distributed in general to the orbital motion
of the planets.}. Because the relation Eq.(\ref{JMplanets}) holds for the
conserved quantities, $J$ and $M$, it strongly suggests some fundamental
origin of the planet system.

\begin{figure}[h]
\begin{center}
\begin{tabular}{cc}
\resizebox{120mm}{!}{\includegraphics{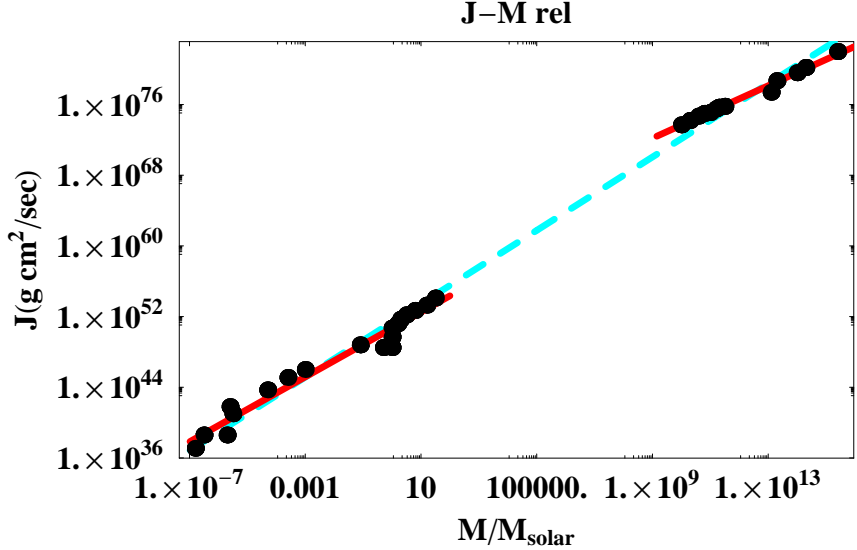}} &  \\ 
& 
\end{tabular}%
\end{center}
\caption{A plot of angular momentum and mass of various astronomical
objects. Deta points includes our planets, stars, galaxies, clusters,...
[1]. The dashed line represents Eq.(\protect\ref{JMplanets}), or more
prcisely $J=1.72\times 10^{-18}M^{2.08}$ in cgs-units. The solid lines
represent local scalings for star series and cluster series. Both the local
inclinations are slightly smaller than the global one.}
\label{fig1}
\end{figure}

Although the extrapolation of the above scaling to the microscopic world
actually holds as the Regge trajectory for Hadrons\footnote{%
This scaling reflects the linear potential for quarks.} $J\simeq \left(
\hbar /\text{Gev}^{2}\right) M^{2}$, it would be more interesting to
extrapolate it to much larger scales, such as galaxies and the clusters of
them, which actually holds with the same proportionality constant as Eq.(\ref%
{JMplanets}). This fact is clearly shown by the dashed line in the Figure %
\ref{fig1}, where the scaling $J\propto M^{2}$ extends over 23-digit [1].

Further examination of this figure reveals the existence of two separate
classes of astronomical objects on this scaling line; the \textsl{star series%
} extending roughly from $10^{-7}M_{\odot }$ to $10^{2}M_{\odot }$, and the 
\textsl{cluster series }from $10^{10}M_{\odot }$ to $10^{16}M_{\odot }$.
Although the whole global scaling looks as $J\propto M^{2}$, the local
scaling within each classes is slightly gentle $J\propto M^{1.5}$ to $%
M^{1.8} $. Since this global scaling extending over 23-digits, if being
true, seems to be very difficult to explain from a single mechanism, we
concentrate on the local scaling for cluster series in this paper. The
objects in this series are primarily characterized by the self-gravity. Thus
we now study this scaling relation based on the self-gravity in the
expanding universe.

A considerable number of studies have already been made on the origin of
angular momentum of galaxies. A typical mechanism of the angular momentum
acquisition for galaxies is the tidal torquing (see for example a review
paper [2]). Let us consider a relatively dens region in the expanding
universe. In general, velocity shear on this region in the Eulerian frame
yields a rotation-like motion of this region in the Lagrangian frame. If
this dens region actually becomes isolated from the other region by turning
around from expanding to contracting due to self-gravity, then this dens
region acquires actual angular momentum. However, it is essential for real
acquisition of angular momentum that the flow of matter eventually becomes
multi-streaming and allows the appearance of vorticity.

The acquired angular momentum 
\begin{equation}
\vec{J}=\int_{V}d^{3}x\left( \vec{x}-\vec{x}_{0}\right) \times \vec{v}\left( 
\vec{x}\right) \rho \left( \vec{x}\right)
\end{equation}%
is estimated by using Zel'dovich approximation [3] with the Lagrange
coordinate $\vec{q}$\ as 
\begin{equation}
\vec{x}\left( {\vec{q},t}\right) =\vec{q}-D\left( t\right) \vec{\nabla}_{%
\vec{q}}\psi \left( \vec{q}\right)  \label{Lagrange}
\end{equation}%
where $D\left( t\right) $ is the linear growing factor and $\psi \left( \vec{%
q}\right) $ is the displacement field. Then the angular momentum acquired at
the time of turnaround $t_{\text{TA}}$ becomes 
\begin{equation}
\vec{J}=-\rho _{0}a\left( t\right) ^{5}\int_{V_{\text{com}}}d^{3}q\left( 
\vec{q}-\vec{q}_{0}\right) \times \dot{D}\left( t\right) \vec{\nabla}\psi
\approx t_{\text{TA}}^{1/3}M^{5/3}  \label{Jbytidaltorquing}
\end{equation}%
where $a\left( t\right) $ is the scale factor, and the integration is taken
over the comoving volume $V_{\text{com}}$. The turn-around time $t_{\text{TA}%
}$\ depends on the mass $M$ and the initial condition for the density
fluctuations.

This formula of the angular momentum Eq.(\ref{Jbytidaltorquing}) is
numerically checked by the high-resolution galaxy-formation simulation using
the P3MSPH code [4]. They have confirmed the power law scaling in Mass $%
M^{5/3}$ in Eq.(\ref{Jbytidaltorquing}), but the time dependence $t_{\text{TA%
}}^{1/3}$ has not been firmly established. According to their calculations,
the angular momentum $J$ of each cluster evolves almost linearly in time
until the turn-around time $t_{\text{TA}}$, when the most of the full $J$ of
the cluster is achieved. Then after the shell crossing time, just after $t_{%
\text{TA}}$, $J$ fluctuates due to the strong and repeated tidal
interactions with neighboring clusters. It is remarkable that the final
angular momentum acquired by each cluster still obeys the scaling law in Eq.(%
\ref{Jbytidaltorquing}) even after such strong disturbances.

Although the amount of angular momentum is mostly acquired before the
turn-around time or the shell crossing time, strongly interacting nonlinear
period after that should washout the previous correlations in general.
Therefore it would be interesting to examine the dynamics in such nonlinear
regime, which may be essential for the establishment of the scaling
relations or any correlations which we now observe.

Moreover, within this cluster series, many other scaling relations have been
reported, such as the increasing velocity dispersion $\sigma $ with the
linear scale $r$, $\sigma ^{2}\propto r$. [5], the scaling in average
mass-density and the size of the astronomical objects $\rho \propto r^{-1.75}
$ [6], power law two-point correlation functions for galaxies $\xi
(r)\approx 20\left( {r/}\text{{Mpc}}\right) ^{-1.8}$ and for clusters of
galaxies $\xi (r)\approx 300\left( {r/}\text{{Mpc}}\right) ^{-1.8}$ [3].
These properties appear to have no relation with the angular momentum
especially in the larger scales. Since these properties hold for non-linear
structures, it is apparent that we cannot attribute all of the origin of
them to the initial conditions. We would like to understand these scaling
properties from a simple principle which will hold in the non-linear stage.
Is there any basic dynamics which realize all such scaling relations?

For this end, we would like to focus on the turbulent motion of the cold
dark matter, i.e. dark turbulence. Since the cold dark matter is
collision-less, we cannot apply the standard view of the baryonic
turbulence. Actually the collision-less particles cannot form local fluid
element which is usually required for fluid description. Thus the dark
turbulence is quite different from the ordinary one and even the eddy
structure or vector perturbation modes may not dominate. We use the word
(dark) turbulence in this sense emphasizing the distinction from the
baryonic turbulence. This point will be especially important to understand
the difference between our dark turbulence from the old paradigm of the
turbulent origin of galaxies. Moreover in our calculations, we never use the
eddy picture of turbulence nor assume the vector mode dominance in
fluctuations.

The purpose of this paper is to consider all of the above scaling relations
from a simple principle. The following is the steps for this purpose. In the
next section 2, we explore the possible origin of the scaling in analogy of
the fluid turbulence, especially in the form of dark matter, which is not
directly rejected from the present observational data. In section 3, we
apply the method of Smoluchowski coagulation equation for Navier-Stokes
equation and derive the Kolmogorov scaling laws, in appropriate form in use
for SGS. In section 4, we explore how extent our proposal can be tested from
astronomical observations. In the last section 5, we summarize our study and
discuss on further extensions and possible applications of the present
discussions.

\section{Fluid description for self-gravitating systems (SGS) and the dark
turbulence}

In the present study on the scaling relations, we have restricted our
considerations to the cluster series of objects such as globular clusters,
galaxies and clusters of galaxies. It is apparent that these structures are
mainly governed by the gravity which has no intrinsic scale. However, this
fact itself does not determine the scaling index. On the other hand, the
virial theorem is expected to hold in the dynamically relaxed systems, and
yields a relation between the velocity dispersion $\sigma $, the linear
scale $r$ and the mass $M$ as $\sigma ^{2}=GM/r$. Recently we have developed
an elaboration of this virial theorem and proposed a concept of the \textsl{%
local virial relation}, which claims that the virial relation holds even a
part of a system, which is not necessarily isolated from the rest of the
system [7], [8]. However, this virial relation is not enough and yet another
relation is necessary to determine the scaling indices. Further, since the
above objects are highly non-linear, the linear perturbation method based on
the primordial density fluctuations is useless. We directly need to consider
highly non-linear regime for our purpose.

We now start our discussion from the reconsideration of the basic fluid
description to analyze SGS. Actually there are many common properties
between the fluid and SGS. Standard fluid is described by the Navier-Stokes
equation for the velocity $\vec{v}\left( \vec{x},t\right) $ 
\begin{equation}
\frac{\partial \vec{v}}{\partial t}+\left( \vec{v}\cdot \vec{\nabla}\right) 
\vec{v}=-\frac{\vec{\nabla}p}{\rho }+\nu \Delta \vec{v},
\label{Navier-Stokes}
\end{equation}%
where $p$ is the pressure, $\rho $ is the mass density, and $\nu $ is the
friction coefficient. This equation admits the following scaling, with
arbitrary real number $\alpha $, 
\begin{eqnarray}
r^{\prime } &=&\lambda ^{-1}r,\;\vec{v}^{\prime }=\lambda ^{-\alpha /3}\vec{v%
},\;  \nonumber \\
t^{\prime } &=&\lambda ^{-1+\left( \alpha /3\right) }t,\;p^{\prime }/\rho
^{\prime }=\lambda ^{-2\alpha /3}p/\rho  \label{scaling}
\end{eqnarray}%
in the inertial regime, \textsl{i.e.} the dissipative term is negligible
compared with the inertial term. Similarly, a set of Jeans equation for SGS
and Poisson equation becomes

\begin{equation}
\begin{array}{c}
\frac{\partial \rho }{\partial t}+\vec{\nabla}\cdot \left( \rho \vec{v}%
\right) =0, \\ 
\frac{\partial \vec{v}}{\partial t}+(\vec{v}\cdot \vec{\nabla})\vec{v}+\vec{%
\nabla}\Phi =0, \\ 
\vec{\nabla}^{2}\Phi =4\pi G\rho ,%
\end{array}
\label{Jeans}
\end{equation}
which also admits the same scaling Eq.(\ref{scaling}) supplemented with $%
\Phi ^{\prime }=\lambda ^{-2\alpha /3}\Phi ,\rho ^{\prime }=\lambda
^{-\left( 2\alpha /3\right) +2}\rho $.

There have been many analytical attempts to reveal the semi-nonlinear
regime, just beyond the linear regime. One of them is the Lagrangian method
such as the Zel'dovich approximation [3]. This is based on the Lagrangian
coordinate $\vec{q}$, which is related to the Eulerian one $\vec{x}$\ as Eq.(%
\ref{Lagrange}), and assumes that the particles develops in the initial
potential. Then the mass-density $\rho \left( {\vec{x},t}\right) $ is simply
given by 
\begin{equation}
\rho \left( {\vec{x},t}\right) =\rho _{0}\det \left( {\delta _{ij}-D\left(
t\right) \frac{\partial ^{2}\psi }{\partial q_{i}\partial q_{j}}}\right)
^{-1}.  \label{ZA}
\end{equation}%
This approximation works fairly well to describe the semi-nonlinear
evolution until inevitable shell-crossing, or caustic surface, appears when
the density $\rho \left( {\vec{x},t}\right) $\ diverges due to the increase
of the {linear growth }factor ${D\left( t\right) }$. Many authors have tried
to avoid or to improve this shortcoming in this approximation. Further,
various sophisticated perturbation methods have been proposed to improve the
accuracy of approximations but this shell-crossing couldn't overcome within
real values [9]. Actually, shell-crossing is not simply a limitation of the
analytic approximations but the initiation of the multi-valued stream, or
the appearance of the family of crossing trajectories [10]. It is important
to notice that the intrinsic non-linear regime starts from this
shell-crossing time, after that the trajectories become chaotic due to the
velocity dispersions and the whole system becomes turbulent\footnote{%
As we have already explained, we simply use the word turbulence for a
general fluid in which each parts of the system is violently mixed together.}%
. Therefore this turbulence should be the clue to solve our problem [11].

At this point, we would like to consider the time scale necessary for the
turbulence to become effective. In general in the standard CDM model,
smaller the linear scale later the shell-crossing time. Therefore the
turbulence is established first at the small scale, say $10^{7}M_{\odot }$,
and then the turbulent region gradually extends up to the larger scales of
galaxy clusters at redshift of order one. On the other hand, the system
acquires most of the angular momentum and kinetic energy which is comparable
to the final amount at the time of turn around, as we have already seen in
the estimation Eq.(\ref{Jbytidaltorquing}) and the demonstration by
numerical calculations [4]. Just after this turn around time, the flow
becomes turbulent at the shell-crossing time. Thus the turbulent velocity
field has been already developed at a given redshift corresponds to its
shell-crossing time. However, further time period will be necessary for this
turbulent flow to become effective and to establish any order such as
scaling relations. This period will be estimated later in section 4, using a
relevant parameter obtained from the observations.

There have been many studies on the fluid description for SGS in the long
history of literature. It is particularly interesting that some authors had
already proposed the galaxy formation due to the turbulent motion [12],
[13]. However, the matter associated with this turbulent motion was assumed
to be baryon, which could easily destroy the observed isotropy of the cosmic
background radiation (CMB) and furthermore the turbulent motion would easily
decay due to the dissipation through the electromagnetic interactions.
Actually the time when this line of study declined overlaps the time when
the precision observation on CMB had been developing.

Now the basic understanding on the cosmic matter contents has been
drastically changed. The dominant substance is not baryon but some
unidentified object which is thought to be collisionless non-relativistic
particles. This is often called as dark matter (DM) emphasizing the absence
of the interaction with baryons, except gravitational interaction. Therefore
what we should consider will be the turbulent motion of DM; thus the title
of this paper cosmic dark turbulence (CDT) comes out. Since DM never destroy
the isotropy of CMB and never dissipate, the turbulent motion of it seems
plausible.

There are several other facts which are crucial for the idea of CDT. (1)
Although the vorticity possesses only the decaying mode $\omega \propto
t^{-1/3}$ within the linear perturbation theory, it can grow in the
non-linear regime [14], [15]. (2) Since the trajectory becomes
multi-streaming, the Kelvin's theorem, which claims the conservation of
vorticity along the trajectory in the non-dissipative flow, becomes useless.
(3) An analogue of the Reynolds number, i.e. the ratio of the dissipation
and the inertial terms, can be calculated in SGS although the system is
time-reversible in the microscopic level. An analogue of viscosity may enter
the system through the dynamical friction $\nu =O\left( 1\right) \ln N\sqrt{%
rGM}/N$, where $N$ is the total particle number. Then the application of the
virial theorem $v^{2}=GM/r$ yields the Reynolds number $R=vr/\nu =O\left(
1\right) N/\ln N$, which turns out to be quite large. However, it must be
kept in mind that such microscopic friction often becomes irrelevant and is
dominated by an effective friction known as eddy viscosity. (4) In general,
the stationary turbulence structure is maintained by a steady energy flow,
on average, from the large to small scales through the cascade of eddies.
This flow is determined by the balance of the energy input at the largest
scale and the dissipation at the smallest Kolmogorov scale. In the case of
SGS, there are many evidences of hierarchical structure formation in which
the coalescence of small clusters successively yield larger clusters. Then
the energy flow should exist but the direction may be opposite; from small
to large. It is important to notice that the essential properties of the
turbulence such as the Kolmogorov laws hold irrespective of the direction of
energy flow. Although the above mentioned dynamical friction may not control
the flow, the energy flow seems to be essential for the scaling property we
consider, irrespective of its direction. This point will be further
discussed in the next section. (4) The most important property of CDT is the
non-linear interaction among various modes from the smallest to the largest,
and the eddy structure is not necessary in our case. This will be well
realized in the self gravitating systems.

\section{Smoluchowski equation and Kolmogorov law}

In order to describe the dynamics of coagulation process in general, the
mean field approach has been widely used and quite successful. We first
consider a simple model in which the whole system is composed from clusters
of various masses, which we measure in some arbitrary unit $m_{0}$. Then the
variable ${n_{i}}\left( t\right) $ is the number of clusters of mass $%
i\times m_{0}$ at time $t.$ A typical evolution equation for ${n_{i}}\left(
t\right) $, for positive integers $i,j,k$, is given by the Smoluchowski
coagulation equation 
\begin{equation}
\frac{dn_{i}}{dt}=\frac{1}{2}\sum\limits_{j+k=i}{K_{j,k}n_{j}n_{k}-}%
n_{i}\sum\limits_{j}{K_{i,j}n_{j},}  \label{smoluchowski}
\end{equation}%
where the kernel ${K_{j,k}}$ is the probability that the clusters $j$ and $k$
coalesce into one [16]. Only several exact solutions are known despite its
simple appearance [17].

We now consider the scaling transformation 
\begin{equation}
i\rightarrow \lambda i
\end{equation}
and look for the solution, of Eq.(\ref{smoluchowski}), which transforms 
\begin{equation}
n_{i}\rightarrow \lambda ^{\mu }n_{i}
\end{equation}
under the above scaling. We suppose that the kernel has a power law
dependence\footnote{%
This kind of kernel often appears in SGS where gravity has no intrinsic
scale.}, $K_{i,j}\propto \left( {ij}\right) ^{\nu /2}$, and also suppose
that the mass flow from small to large scales per cluster $\dot{n}%
_{i}i/n_{i} $ is a constant. Then, since $\dot{n}_{i}$ in Eq.(\ref%
{smoluchowski}) scales as $\dot{n}_{i}\rightarrow \lambda ^{\nu +2\mu }\dot{n%
}_{i}$ (and the time scales as $t\rightarrow \lambda ^{-\nu -\mu }t$), we
have $1+\nu +\mu =0$. Thus, $\mu =-1-\nu ,$ and therefore 
\begin{equation}
n_{i}=ci^{-\nu -1},  \label{scaling-n}
\end{equation}
where $c$ is a mass $i$-independent constant.

It is important to notice that this is not a genuine stationary state since $%
\dot{n}_{i}$ does not vanish and the configuration $n_{i}$ is always
changing. It will be still possible to add an appropriate source term, which
is non-zero only for small values of $i$, to maintain the local balance $%
\dot{n}_{i}=0$ for each cluster $i$, while keeping the same index $-\nu -1$
in Eq.(\ref{scaling-n}). However, as is shown in the above, this index is
determined mainly by the actual steady flow and the non-linear kernel ${%
K_{j,k}}$ which controls it. If we supposed the local balance $\dot{n}_{i}=0$
instead, we would have no robust result which is free from a choice of the
source term\footnote{%
The problem why the mass flow per cluster should be constant still remains
though Eq.(\ref{scaling-n}) successfully describes many simulations for
Smoluchowski equation.}.

This type of equation Eq. (\ref{smoluchowski}) appears in various fields of
physics. Actually, the Fourier transform of Eq.(\ref{Navier-Stokes}) becomes 
\begin{equation}
\frac{\partial v_{\vec{k}}^{\alpha }}{\partial t}=-ik_{\beta }\sum\limits_{%
\vec{p}+\vec{q}=\vec{k}}{K_{\vec{k}\mathbf{,}\alpha \gamma }v_{\vec{p}%
}^{\beta }v_{\vec{q}}^{\gamma }-\nu k}^{2}v_{\vec{k}}^{\alpha }+\text{source
term}  \label{FTNS}
\end{equation}
where 
\begin{equation}
{K_{\vec{k}\mathbf{,}\alpha \gamma }=\delta }_{\alpha \gamma }-\frac{%
k_{\alpha }k_{\gamma }}{k^{2}},
\end{equation}
\begin{equation}
v^{\alpha }\left( \overrightarrow{x},t\right) {=}\int \frac{d^{3}k}{\left(
2\pi \right) ^{3}}e^{i\overrightarrow{k}\overrightarrow{x}}v{_{\vec{k}%
}^{\alpha }}\left( t\right) {,}  \label{FTu}
\end{equation}
\begin{equation}
\sum\limits_{\vec{p}+\vec{q}=\vec{k}}=\int d^{3}pd^{3}q\delta ^{3}\left( 
\vec{p}+\vec{q}-\vec{k}\right) .
\end{equation}
The analogy to Eq.(\ref{smoluchowski}) is apparent; the Fourier mode $v_{%
\vec{k}}^{\alpha }\left( t\right) $ replaces the number of clusters $%
n_{i}\left( t\right) $. Furthermore, the analogy to obtain the power law
distribution Eq.(\ref{scaling-n}) is possible.

We consider the scaling transformation 
\begin{equation}
\vec{k}\rightarrow \lambda \vec{k}\mathbf{,}\text{ or }\vec{x}\rightarrow
\lambda ^{-1}\vec{x}  \label{timeslambda}
\end{equation}%
and we look for the solution, for Eq.(\ref{FTNS}), which transforms 
\begin{equation}
v_{\vec{k}}^{\alpha }\rightarrow \lambda ^{\mu }v_{\vec{k}}^{\alpha }
\label{vscale}
\end{equation}%
under the above scaling. Here we adopt the random phase approximation, in
which we progressively take in account the phases of Fourier transformed
quantities from lower to higher orders. Therefore, after the inverse-Fourier
transformation, they do not simply reproduce the original velocity field in
real space, but produce instead statistical variables. We assume the energy
flow is a constant $dE\left( \overrightarrow{x}\right) /dt=$ $\varepsilon $,
in analogy with the case for Smoluchowski equation. Since the kernel scales
as $\left\vert -ik_{\beta }\sum\limits_{\vec{p}+\vec{q}=\vec{k}}{K_{\vec{k}%
\mathbf{,}\alpha \gamma }}\right\vert \rightarrow \lambda ^{4}\left\vert
-ik_{\beta }\sum\limits_{\vec{p}+\vec{q}=\vec{k}}{K_{\vec{k}\mathbf{,}\alpha
\gamma }}\right\vert $, we have the scaling 
\begin{equation}
\frac{\partial v_{\vec{k}}^{\alpha }}{\partial t}\rightarrow \lambda
^{4+2\mu }\frac{\partial v_{\vec{k}}^{\alpha }}{\partial t}
\end{equation}%
where we consider the inertial regime in which the dissipation and the
source terms do not dominate. This scaling suggests that the time should
scale as $t\rightarrow \lambda ^{-4-\mu }t$ for consistency. Then, since $%
E\left( \vec{x}\right) =v\left( \vec{x}\right) ^{2}/2$, $dE\left( \vec{x}%
\right) /dt$ scales as 
\begin{equation}
dE\left( \vec{x}\right) /dt\rightarrow \lambda ^{10+3\mu }dE\left( \vec{x}%
\right) /dt,
\end{equation}%
which must be a constant $\varepsilon $. Then we have 
\begin{equation}
\mu =-\frac{10}{3}  \label{gamma}
\end{equation}%
and , $v^{\alpha }\left( \overrightarrow{x},t\right) ${\ in }Eq.(\ref{FTu})
scales as 
\begin{equation}
v^{\alpha }\left( \overrightarrow{x}\right) \rightarrow \lambda ^{3+\mu }{\ }%
v^{\alpha }\left( \overrightarrow{x}\right) =\lambda ^{-1/3}{\ }v^{\alpha
}\left( \overrightarrow{x}\right) {\ .}
\end{equation}%
Therefore, since $\varepsilon \propto $ $\left( v_{\vec{k}}^{\alpha }\right)
^{3}$ from Eq.(\ref{FTNS}), we have 
\begin{equation}
v^{\alpha }\left( \vec{x}\right) =\left( \left\vert \vec{x}\right\vert
\varepsilon \right) ^{1/3}f\left( \frac{\vec{x}}{\left\vert \vec{x}%
\right\vert }\right)  \label{Kolmogorov1/3}
\end{equation}%
where $f$ is a function of only the direction $\overrightarrow{x}/\left\vert 
\vec{x}\right\vert $.\ As was explained in the above, $v^{\alpha }\left( 
\vec{x}\right) $ is not simply the original velocity field in real space,
but a statistical variable. In other words, the argument $\vec{x}$ in $%
v^{\alpha }\left( \vec{x}\right) $ does not designate any particular
location but only the scale $\left\vert \vec{x}\right\vert $\ is relevant.
Thus $v^{\alpha }\left( \vec{x}\right) $ represents an average velocity
difference or the velocity dispersion at the separation $\left\vert \vec{x}%
\right\vert $. Then Eq.(\ref{Kolmogorov1/3}) corresponds to the Kolmogorov
phenomenological theory [18]. In the same way, the energy per mass $E_{k}$
for the Fourier mode $k\equiv \left\vert \vec{k}\right\vert $ becomes 
\begin{eqnarray*}
E_{k} &\equiv &\int dre^{ikr}E\left( r\right) \\
&=&\int dre^{ikr}v\left( r\right) ^{2}/2=O\left( 1\right) \Gamma \left(
5/3\right) \varepsilon ^{2/3}k^{-5/3}
\end{eqnarray*}%
where $r\equiv \left\vert \vec{x}\right\vert $, i.e.

\begin{equation}
E_{k}=O\left( 1\right) \varepsilon ^{2/3}k^{-5/3}.  \label{Kolmogorov5/3}
\end{equation}
This corresponds to the Kolmogorov 5/3-law which holds in the inertial
regime of turbulent fluid. These exponents are quite robust in the inertial
regime independent from the spatial dimension, how the energy is dissipated
or injected.

Also in this fluid case, the constant energy flow, but not the exact steady
state, has been essential to derive these robust results independent from
the form of energy supply. The most relevant is the non-linear kernel which
drives the constant flow but not simply a local balance between the
dissipation and the energy supply from the source term. This point may be
confusing in the context of the familiar turbulence in laboratory, which
must be supported by steady injection of energy from outside. This
laboratory turbulence will be a special case when the Reynolds number is
relatively small, say $100-1000$. In the huge system, such as the galaxy
whose' Reynolds number $\approx 10^{10}$, which has wide inertial range, we
will be able to check this argument\footnote{%
As in the previous case, the problem why the energy flow should be constant
still remains though the results Eq.(\ref{Kolmogorov1/3}- \ref{Kolmogorov5/3}%
) successfully describe the nature and many simulations.}.

In the same spirit as for the Navier-Stokes equation for fluid, we can argue
the Fourier transform of Eq.(\ref{Jeans}) for SGS, 
\begin{equation}
\frac{\partial v_{\vec{k}}^{\alpha }}{\partial t}=-ik_{\beta }\sum\limits_{%
\vec{p}+\vec{q}=\vec{k}}{K_{\vec{k}\mathbf{,}\alpha \gamma }v_{\vec{p}%
}^{\beta }v_{\vec{q}}^{\gamma }-\nu k}^{2}v_{\vec{k}}^{\alpha }+i\frac{%
k^{\alpha }}{k^{2}}4\pi G\delta _{\vec{k}}  \label{FTJeans2}
\end{equation}%
\bigskip which is also similar to the Smoluchowski equation, where 
\begin{equation}
\delta _{\vec{k}}\left( t\right) =\int d^{3}xe^{i\overrightarrow{k}%
\overrightarrow{x}}\rho \left( \vec{x},t\right) {,}  \label{FTdelta}
\end{equation}%
and $\nu =O\left( 1\right) \ln N\sqrt{rGM}/N$. Poisson equation Eq.(\ref%
{Jeans}) is used for the above expression for the source term. The
continuity equation becomes 
\begin{equation}
\dot{\delta}_{\vec{k}}-ik^{\alpha }\sum\limits_{\vec{p}+\vec{q}=\vec{k}}v{_{%
\vec{p}}^{\alpha }\delta _{\vec{q}}=0.}  \label{FTJeans1}
\end{equation}%
We now look for the solution, for Eqs.(\ref{FTJeans2}, \ref{FTJeans1}),
which transforms like Eq.(\ref{vscale}). The essential structure of the
equations is the same as in the fluid case, except the explicit source term
from the density fluctuations $\delta _{\vec{k}}$. We suppose it scales as 
\begin{equation}
\delta _{\vec{k}}\rightarrow \lambda ^{\nu }\delta _{\vec{k}}  \label{delta}
\end{equation}%
under the scaling transformation Eq.(\ref{timeslambda}). All the exponents
are consistently determined as before, in the inertial regime, and we have
Eq.(\ref{gamma}) and 
\begin{equation}
\nu =5+2\mu =-\frac{5}{3}.  \label{eta}
\end{equation}%
Thus the Kolmogorov laws Eqs.(\ref{Kolmogorov1/3})-(\ref{Kolmogorov5/3})
still hold in the inertial regime.

In the case of collision-less SGS, the dissipation due to the binary
collision nor the external driving force are not thought to control the
steady energy flow. However in case of the cold dark matter, we naturally
expect a hierarchical coalescence process to form structures. Since smaller
clusters coalesce to form a larger cluster, the energy flows from small
scale to large scale in real space. This is more like the process described
by Smoluchowski coalescence equation. Although the flow direction is
opposite to the case of fluid turbulence, in which the energy flows from
large to small, the existence of a steady energy flow is essential to derive
Kolmogorov scaling relations. Thus we could obtain the same Kolmogorov laws
in SGS. It is also important in the above calculations that we never used
the eddy structure. This demonstrates our claim that the CDT does not
require eddies or vector mode dominance.

The above Kolmogorov equations provide us a new relation which is essential
to solve our problem. The velocity difference at the scale $\left\vert 
\overrightarrow{x}\right\vert :v^{\alpha }\left( \overrightarrow{x}\right) $
in Eq.(\ref{Kolmogorov1/3}) is a statistical variable which can also be
understood as the velocity dispersion $\sigma _{r}$ at the scale $%
r=\left\vert \overrightarrow{x}\right\vert $. Therefore our first relation
is 
\begin{equation}
\sigma _{r}=\left( r\varepsilon \right) ^{1/3},  \label{sigma at r}
\end{equation}%
which claims that the velocity dispersion increases with the scale. A
numerical factor of $O\left( 1\right) $ on the right hand side of Eq.(\ref%
{Kolmogorov1/3}) is absorbed into the parameter $\varepsilon $. By utilizing
the local virial relation with this equation, we have the mass at the scale $%
r$ as 
\begin{equation}
M_{r}=\frac{r\sigma _{r}^{2}}{G}=\frac{\varepsilon ^{2/3}}{G}r^{5/3},
\label{M at r}
\end{equation}%
or equivalently, the mass density at the scale $r$ as 
\begin{equation}
\rho _{r}=\frac{M_{r}}{\left( 4\pi /3\right) r^{3}}=\frac{\varepsilon ^{2/3}%
}{\left( 4\pi /3\right) G}r^{-4/3},  \label{rho at r}
\end{equation}%
which claims that the mass density reduces with the scale. This is fully
consistent with Eqs.(\ref{eta}, \ref{delta} and \ref{FTdelta}), as should
be. We now examine possible observational tests on the above results in the
next section.

\section{\textbf{Observational tests}}

We now examine the scaling relations Eqs.(\ref{sigma at r})-(\ref{rho at r}%
), which we obtained in the previous section, by applying them to the
observational data recently available. Certainly our discussion based on the
autonomous dynamics of CDT will be limited both from upper and lower scales.
The upper boundary comes from the fact that the larger scale structures are
better described by the deterministic evolution which is often analyzed by
linear approximations. The lower boundary comes from the fact that the
smaller scale structures have nothing to do with the dark matter turbulence
nor hierarchical coalescence processes but may be better described by the
baryonic interactions or the hierarchical decay processes. We should be
careful for such applicability range in the following tests. A single
parameter $\varepsilon $, the steady energy flow per particle, in our
argument on CDT, will be found below.

\subsection{Scale dependent velocity dispersion}

\begin{figure}[h]
\begin{center}
\begin{tabular}{cc}
\resizebox{120mm}{!}{\includegraphics{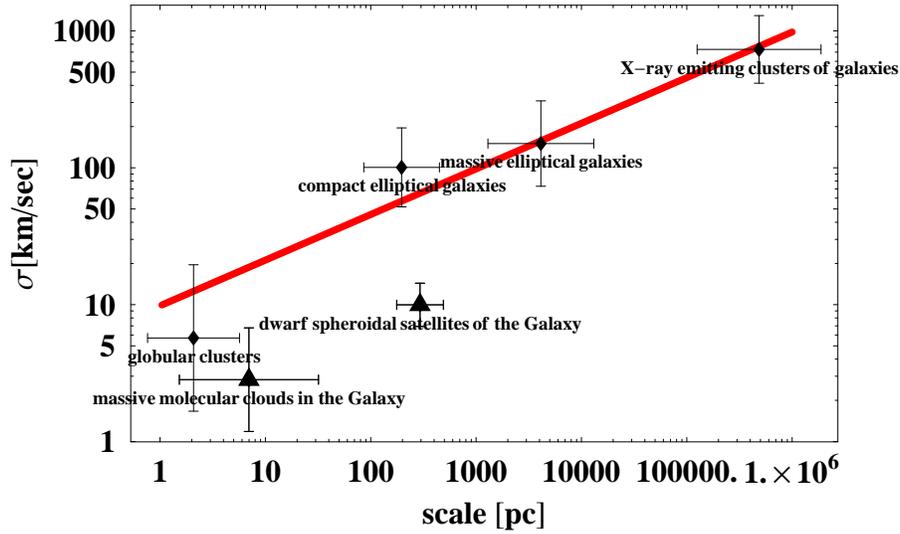}} &  \\ 
& 
\end{tabular}%
\end{center}
\caption{The line-of-sight velocity dispersion for various astronomical
objects as a function of the linear scale. The deta points are from Sanders
\& McGaugh [5], and represents, from large scale to small scale, X-ray
emitting clusters of galaxies, massive elliptical galaxies, dwarf spheroidal
satellites of the Galaxy, compact elliptical galaxies, massive molecular
clouds in the Galaxy, and globular clusters. The solid line representing Eq.(%
\protect\ref{sigma at r}) is drawn disregarding the data of satellites and
molecular clouds (marked as filled triangles) since they might be
dynamically contaminated from our Galaxy or from the Galactic disk.}
\label{fig2}
\end{figure}

We first apply Eq.(\ref{sigma at r}) for observations for galaxies and
clusters. The line-of-sight velocity dispersion for various astronomical
objects as a function of the linear scale is shown in the Figure \ref{fig2}.
The data points are from Sanders \& McGaugh [5], and represents, from large
to small scales, X-ray emitting clusters of galaxies, massive elliptical
galaxies, dwarf spheroidal satellites of the Galaxy, compact elliptical
galaxies, massive molecular clouds in the Galaxy, and globular clusters.
Among them, dwarf spheroidal satellites and molecular clouds (marked as
filled triangles) are irrelevant for our argument. This is because the
extended dwarf spheroidal satellites of the Galaxy are located near to the
dominant Galaxy, and therefore any systematic tidal effects cannot be
avoided [19]. On the other hand, because the massive molecular clouds in the
Galaxy are located very inside the galaxy disk, any dynamical contamination
cannot be avoided [20].

On top of the observational data, disregarding the above mentioned
irrelevant species, a solid line is drawn which represents Eq.(\ref{sigma at
r}) with $\varepsilon =0.3[$cm$^{2}/\sec ^{3}]$. This value will be fixed
throughout this paper. It is apparent from the graph that the larger-scale
structure has higher velocity dispersion. It is important to notice that
this scaling law only holds globally in the whole scale region, from $1pc$
to $1Mpc$, but not locally within the individual species of objects. For
example, we cannot claim the scaling only from the data for galaxies because
of huge scatter. The above value for the parameter $\varepsilon $ allows
error of factor two when we consider the global fitting.

This value $\varepsilon =0.3[$cm$^{2}/\sec ^{3}]$ also determines the
intrinsic upper limit of the size for the applicability of our argument. The
energy of an object of size $r$ is roughly given by $\sigma ^{2}=\left(
\varepsilon r\right) ^{2/3}$ per mass, from Eq.(\ref{sigma at r}), which
should be accumulated by the steady energy flow during the time span $T$ as $%
\varepsilon T$. Equating them $r=\varepsilon ^{1/2}T^{3/2}$ and putting the
cosmic age for $T$, we have $r\approx 50Mpc$, which yields the intrinsic
upper limit in size. Another estimation is also possible if we temporary
assume the eddy picture of turbulence. A necessary time for an eddy of size $%
r$ to make one revolution is $T=2\pi r/v$ with Eq.(\ref{sigma at r}).
Requiring that this should not exceed the cosmic age for the effective
relaxation, makes the upper limit as $r=\left( 2\pi \right)
^{-3/2}\varepsilon ^{1/2}\left( \cos \text{mic age}\right) ^{3/2}\approx $ $%
3.2Mpc$. Thus we roughly understand that our discussion cannot be applied
beyond these scales.

\subsection{Mass-Luminosity ratio}

\begin{figure}[h]
\begin{center}
\begin{tabular}{cc}
\resizebox{120mm}{!}{\includegraphics{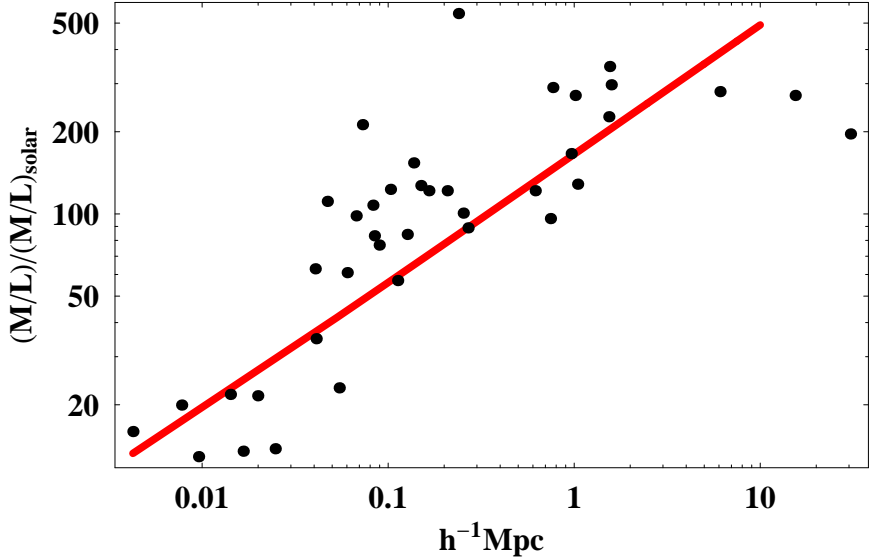}} &  \\ 
& 
\end{tabular}%
\end{center}
\caption{Mass-Luminosity ratio in various scales. The points represent
various astronomical objects [21]. The solid line represents Eq.(\protect\ref%
{M/L}) with $\protect\varepsilon =0.3[$cm$^{2}/\sec ^{3}]$ and $\ \protect%
\rho _{LM}/\protect\rho _{DM}=6\times 10^{-4} $ at $r=1$AU. }
\label{fig3}
\end{figure}

We now examine the mass density as a function of size. From Eq.(\ref{rho at
r}), we have 
\begin{equation}
\rho _{DM}=\frac{\varepsilon ^{2/3}}{\left( 4\pi /3\right) G}r^{-4/3},
\label{rhoDM}
\end{equation}%
for dark matter. On the other hand, many observations indicate that the
luminous matter behaves differently, 
\begin{equation}
\rho _{LM}=\rho _{LM0}r^{-1.8}  \label{rhoLM}
\end{equation}%
where $\rho _{LM0}$ is some constant. Combining these equations of different
nature, the ratio of the gravitational mass and luminous mass becomes 
\begin{equation}
M/L\equiv \frac{\rho _{DM}+\rho _{LM}}{\rho _{LM}}\propto r^{0.48},
\label{M/L}
\end{equation}%
which claims that the ratio $M/L$ increases with scale. This is compared
with observations of Bahcall et al. [21]. The solid line in Figure \ref{fig3}
represents Eq.(\ref{M/L}), with $\varepsilon =0.3[$cm$^{2}/\sec ^{3}]$,
which is always fixed in this paper, and $\ \rho _{LM}/\rho _{DM}=6\times
10^{-4}$ at $r=1$AU. The latter parameter is just for a calibration and is
not at all any prediction of $\rho _{LM}/\rho _{DM}$ at $1AU$, since the DM
turbulence cannot affect the physics at $1AU$. Allowing large scatter, the
solid line traces the overall tendency of the data up to about $10Mpc$,
beyond this scale there seems to be a systematic saturation of the ratio. We
must notice that CDT only predicts Eq.(\ref{rhoDM}) for DM, but Eq.(\ref%
{rhoLM}) is a result of observations.

\subsection{Mass-angular momentum relation}

We now study the angular momentum-mass relation, which has been our original
motivation of the present study. By utilizing the rigid body approximation,
virial relation, and Eq.(\ref{M at r}), we have the expression 
\begin{equation}
\frac{J}{M^{2}}=\frac{2}{5}G^{4/5}\varepsilon ^{-1/5}M^{-1/5}  \label{J/M^2}
\end{equation}%
with $\varepsilon =0.3$ as before. It is worth mentioning that this mass
dependence of the angular momentum $J\propto M^{9/5}$ is very similar to the
one predicted from the tidal torquing Eq.(\ref{Jbytidaltorquing}) $J\propto
M^{5/3}$. A mild dependence on $M$ in the factor $t_{\text{TA}}^{1/3}$ in
Eq.(\ref{Jbytidaltorquing}) even works to reduce the difference of them. We
would like to emphasize that the latter is the angular momentum acquired by
the turn around time and the former is one regulated by turbulence later.

\begin{figure}[h]
\begin{center}
\begin{tabular}{cc}
\resizebox{120mm}{!}{\includegraphics{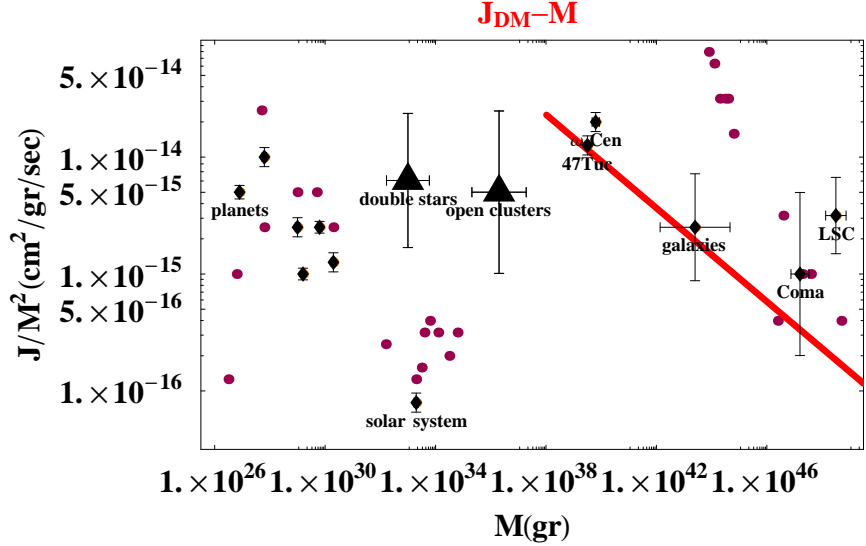}} &  \\ 
& 
\end{tabular}%
\end{center}
\caption{The relation between $J/M^{2}$ and $M$ for various objects; simple
points are from Muradian et al. [1], and points with error bars are from
Brosche [22]; Brosche et al. [23]. Double stars and the open clusters,
marked by triangles, are not considered to be SGS and thus excluded from our
discussion. The solid line represents our relation Eq.(\protect\ref{J/M^2})
with $\protect\varepsilon =0.3$.}
\label{fig4}
\end{figure}

This result, expressed as the solid line in Figure \ref{fig4}, is compared
with observations; simple points are from Muradian et al. [1], and points
with error bar are from Brosche [22]; Brosche \& Tassie [23]. It should be
noticed that the vertical axis $J/M^{2}$ extends over 2-digit, while the
horizontal axis $M$ extends over 20-digit. Within this extension, the range
of our consideration is only the right half of this graph for cluster
series. \ Although the data have huge scatter, the relation Eq.(\ref{J/M^2})
is not excluded. It is also noticed that in the left half of this graph, a
similar line with the same inclination can fit the rest of the data which
includes planets and stars. Let us have discussion on this fact in the
summary section.

\subsection{Power spectrum of density fluctuations $P\left( k\right) $}

The power spectrum of the density fluctuations $P\left( k\right) $ can be
expressed in terms of the two-point correlation function $\xi \left(
r\right) $ as 
\begin{equation}
P\left( k\right) =4\pi \int_{0}^{\infty }r^{2}dr\frac{\sin \left( kr\right) 
}{kr}\xi \left( r\right) ,  \label{P(k)}
\end{equation}
in which $\xi \left( r\right) $ can be expressed by the density function $%
\rho \left( r\right) $ which is already obtained as Eq.(\ref{rho at r}) for
DM. Thus, we have 
\begin{equation}
\xi \left( r\right) =\frac{\rho \left( r\right) }{\rho _{0}}=\xi
_{0}r^{-4/3},
\end{equation}
where $\xi _{0}=$ $\varepsilon ^{2/3}/\left( 4\pi G\rho _{0}/3\right) $, and 
\begin{equation}
P\left( k\right) =2\sqrt{3}\pi \Gamma \left( \frac{2}{3}\right) \xi
_{0}k^{-5/3}.  \label{P(k)estimated}
\end{equation}

\begin{figure}[h]
\begin{center}
\begin{tabular}{cc}
\resizebox{120mm}{!}{\includegraphics{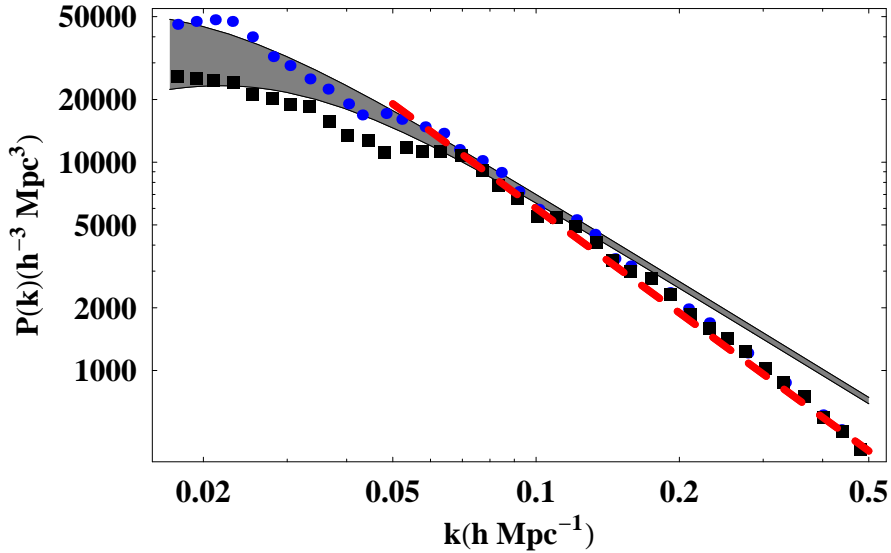}} &  \\ 
& 
\end{tabular}%
\end{center}
\caption{The power spectrum of density fluctuations. The points and squares
represent the data from 2dF and SDSS(DR5), respectively [25]. The shaded
region represents the average of the other type of observations [26]. The
dashed line is our result Eq.(\protect\ref{P(k)estimated}) with the
normalization $\protect\sigma _{8}=0.8$ suggested from WMAP data [24]. This
dashed line is an asymptotic extrapolation of the non-linear regime, where
our CDT model is justified, and the partial agreement with 2dF/SDSS(DR5)
results should not be taked seriously. The CDT model cannot describe this
linear to quasi-linear regime with sufficient confidence. }
\label{fig5}
\end{figure}

This relation for DM is expressed as the dashed line in Figure \ref{fig5},
with the appropriate parameter $\rho _{0}$\ for normalization so that it
matches to WMAP data $\sigma _{8}=0.8$ [24]\footnote{%
Estimated $P\left( k\right) $ in Eq.(\ref{P(k)estimated}) becomes
meaningless for smaller $k$ where the scenario of CDT cannot be applied.
Therefore normalization is applied only for $k>0.07$.}. Immediate caution
must be mentioned here. The scale range shown in Fig.(\ref{fig5}) is linear
to quasi-linear regime($0.02<k\left( \text{hMpc}^{-1}\right) <0.5$). The
linear regime is outside of the applicability of our turbulence model, and
the quasi-linear regime is marginal. Therefore the dashed line in the figure
should be understood as an asymptotic extrapolation from the non-linear
region ($0.5<k\left( \text{hMpc}^{-1}\right) $) where CDT model can
describe. Although an applicability of our model is limited, this regime is
important in the sense that the bias, the ratio of the density fluctuations
of DM and luminous matter, becomes very simple and is almost a constant in
this regime. Therefore luminous matter power spectrum obtained from direct
observations faithfully traces DM power spectrum predicted from our CDT
model. Keeping these facts and restrictions in mind, it would be interesting
to compare our results with observations.

The above predictions from CDT model is compared with the recent
observational data of 2dF and SDSS(DR5) [25], which are drawn, respectively,
by dots and squares. The agreement of the slope for the non-linear region $%
k>0.07$ should not be taken seriously as was explained in the above; our CDT
model cannot explain the power spectrum in this regime. On the other hand,
other type of observations is often approximately expressed by an analytic
form [26], 
\begin{equation}
P_{\text{obs}}\left( k\right) =\frac{2\pi ^{2}}{k^{3}}\frac{(k/k_{0})^{1.6}}{%
1+(k/k_{c})^{-2.4}}.  \label{P(k)shaded}
\end{equation}%
This is shown as gray shade in Figure \ref{fig5}, with the parameter $%
k_{0}=0.19(h^{-1}$Mpc$)^{-1}$, and $k_{c}=0.015-0.025(h^{-1}$Mpc$)^{-1}$.

\subsection{Scale dependent Magnetic field}

Now we consider a possible relation between our arguments on CDT and the
large scale distribution of cosmic magnetic fields. Actually, the magnetic
field $\vec{B}$ and the vorticity $\vec{\omega}\equiv rot\vec{v}$\ follow
the very similar equations of motion and a rough argument based on this fact
is found in the book by Landau \& Lifshits [27], section 74. Generation of
the cosmic magnetic field is studied in Ref. [28], and further development
of the magnetic field is discussed in Ref. [29], for example. Magnetic
fields in the higher redshift is discussed in Ref. [30-31].

Although a large number of studies have been made on cosmic magnetic fields,
little is known about the generation and development of them on all scales.
Thus we simply assume here the following conditions as a starting point for
discussions.

\begin{enumerate}
\item The dynamo mechanism is a universal process to convert the kinetic
energy into magnetic fields in various cosmic scales.

\item Baryons should faithfully follow the motion of DM at least at later
stage of the universe.

\item The conversion rate $\Gamma $ is a constant in all scales.
\end{enumerate}

Immediate explanations for them are necessary. The assumption 1 is not at
all verified at present. However if the turbulence is the essence to develop
and regulate the magnetic field, and the similarity in the equations for $%
\vec{B}$ and $\vec{\omega}$\ were true, then such model can be workable. The
assumption 2 may sound strange, because one usually imagine that DM cannot
interact with baryons, it simply form a static potential well, in which
baryons fall straightly. However, since DM is moving according to Eq.(\ref%
{sigma at r}), with velocity dispersion, its potential well may not be
\textquotedblleft static\textquotedblright . Furthermore in the study of
angular momentum of galaxies, there is empirical relation that the DM has
the same specific angular momentum as that for baryons [32]. We have good
grounds for thinking that the baryons are dynamically coupled to turbulent
motions of DM, because the empirical relation is widely supported. The
assumption 3 is total ad hoc; we would like to consider the deviation of the
scaling in the magnetic field, if any, is partially due to the variation of
this conversion rate.

\begin{figure}[h]
\begin{center}
\begin{tabular}{cc}
\resizebox{120mm}{!}{\includegraphics{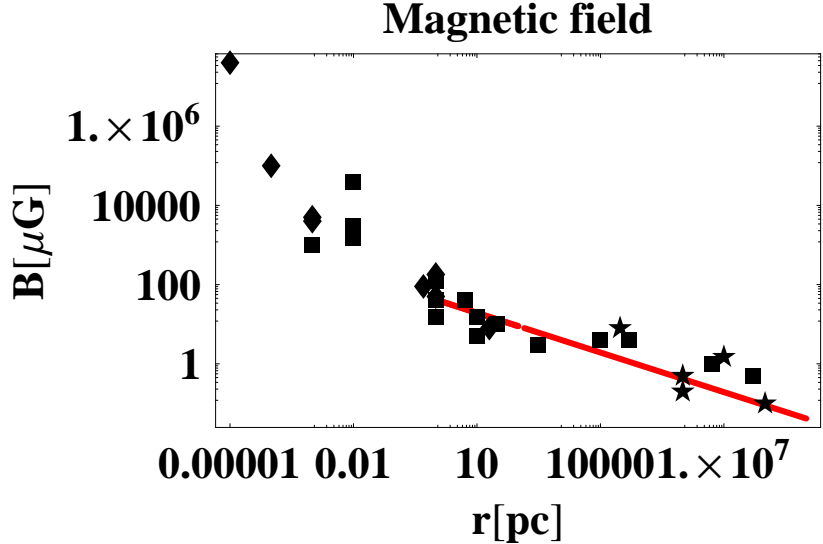}} &  \\ 
& 
\end{tabular}%
\end{center}
\caption{Cosmic magnetic fields in various scales. The deta are from Ref.
[33], marked by stars and Ref. [34], marked by squares and diamonds. The
solid line represents Eq.(\protect\ref{Bestimated}) with parameter $\Gamma
=0.02$. This fit is violated below the distance about the parsec scale, and
another power seems to develope below there. }
\label{fig6}
\end{figure}

If we adopt the above conditions,then the resultant magnetic field should
have a systematic scale dependence which reflects the scale dependent
kinetic energy in the turbulence. The kinetic energy at the scale $r$
becomes 
\begin{equation}
\frac{1}{2}\rho _{DM}\left\langle {v^{2}}\right\rangle =\frac{3}{8\pi G}%
\varepsilon ^{4/3}r^{-2/3}.
\end{equation}%
Then, since the portion of $\Gamma $ of this quantity is supposed to tern
into the energy density of the magnetic field $\vec{B}^{2}/\left( 8\pi
\right) $,\ we can estimate the magnetic field produced from the dynamo
mechanism as 
\begin{equation}
\vec{B}_{estimated}=\left( \frac{3\Gamma }{G}\right) ^{1/2}\varepsilon
^{2/3}r^{-1/3}.  \label{Bestimated}
\end{equation}%
On the other hand, the strength of the magnetic field $B$ obtained from
cosmological observations in various scales are depicted in Figure \ref{fig6}
[33], [34]. Guided these data, we have chosen the value $\Gamma =0.02$,\
which means that the energy of the magnetic field is 12 percent of the
kinetic energy in the baryonic form. The estimated magnetic field Eq. (\ref%
{Bestimated}) from CDT, with this $\Gamma $, is depicted in Figure \ref{fig6}
by the solid line. Reflecting the above bold assumptions, the fit is not so
remarkable though general tendency is not excluded from observations.
Furthermore the magnetic fields seem to form steeper slope at smaller scale
than about parsec, which coincides with the smallest scale of the cluster
series, only beyond that we can apply the argument of CDT.

For the larger scales beyond about $10$ Mpc, the dynamical time becomes
larger than the Hubble time, where we cannot apply our argument.

\section{Summary and Discussions}

We summarize our study and discuss on further extension and possible
applications of it. We started our study from an interesting scaling law%
\footnote{%
We concluded that the index is not $2$ but $1.8$ within the cluster series.
The same index is suggested for star series. See below.} $J\propto M^{2}$
which holds for angular momentum $J$\ and the mass $M$\ of astronomical
objects from the scale of planets toward that of clusters of galaxies.
Focussing on the cluster series, such as globular clusters, galaxies and
clusters of galaxies, we extracted the nature of the fully non-linear stage
of dark matter, after the formation of caustic surface, in analogy between
SGS and the fluid turbulence. The essential Kolmogorov laws were rederived
from the Fourier transformed Navier-Stokes equation and equations for SGS,
by demanding the existence of a steady energy flow characterized by a single
parameter $\varepsilon $. This is an analogue of the case of Smoluchowski
coagulation equation, which admits scaling solution by demanding the steady
mass flow. Then we tried to test our considerations in several cosmological
observations, such as the velocity dispersion, Mass-Luminosity ratio, $J-M$
relation, Power spectrum of density fluctuations, and the cosmic magnetic
fields. They all point the concordant value for the constant energy flow per
mass: $\varepsilon =0.3$cm$^{2}/$sec$^{3}$.

Finally we would like to point out several issues on which we should address
in our subsequent studies for CDT.

\begin{enumerate}
\item We would like to evaluate the energy flow $\varepsilon $ for DM, which
may be associated with the hierarchical coalescence process in which smaller
clusters continuously form larger clusters. This process provides the
bottom-up scenario for the formation of large scale structure especially in
the cold DM model. We would like to check whether the value we used $%
\varepsilon \approx 0.3[$cm$^{2}/\sec ^{3}]$ is consistent with the
hierarchical coalescence evolution of DM. Roughly estimated, this
accumulation rate of kinetic energy yields a galaxy within about $10^{8}$
years.

\item If the above is the case for DM, then we would like to apply the same
analysis to our planet system, which is also considered to have the
coalescence evolution as its origin [35]. Actually in the left half of
Figure \ref{fig3}, the planet and star systems seem to admit the fitting
line with the same slope $-1/5$ as the case for DM but with larger parameter
value for $\varepsilon $.\ It turns out that the appropriate value for $%
\varepsilon $\ becomes of order $10^{15}$, which seems to be too large to be
explained from a simple coalescence process as in the case for DM. Some
violent mechanism, which allow huge energy transfer rate or catastrophic
coalescence, is expected in the case for planet formation process.

\item In the fluid turbulence, there holds another relation often called
Kolmogorov 4/5 law 
\begin{equation}
\left\langle \left( \delta v\left( \vec{r}\right) \right) ^{3}\right\rangle
=-\frac{4}{5}\varepsilon r,
\end{equation}%
which is essential to explain the energy flow takes place from low frequency
modes to high frequency modes on average. Then what is the analogous
relation for SGS and how is it relevant in the universe? Although we do not
know the answer, it will be important to notice the fact that SGS [36]
shares some common properties with the fluid turbulence [37], such as the
negative skewness and the exponential distribution function of the
velocity-difference, etc. These common properties will be a good starting
point for further discussion.

\item It would be interesting if we could actually transform Eq.(\ref%
{FTJeans2}) into the Smoluchowski form like Eq.(\ref{smoluchowski}) which
directly expresses the coalescence evolution of SGS. If this is the case,
the possible scaling solution, which may asymptotically realize, may
correspond to the Schechter function [3] which has a typical form, 
\begin{equation}
\phi \left( L\right) dL=\eta _{\ast }\left( \frac{L}{L_{\ast }}\right)
^{-1.25}\exp \left( -\frac{L}{L_{\ast }}\right) \frac{dL}{L_{\ast }},
\end{equation}%
where $\phi \left( L\right) dL$ represents the frequency of the object of
luminosity $L$. If the cosmological objects are formed after many
coalescence processes, then the Smoluchowski equation will be more
appropriate than the Press-Schechter theory [3], in which a single collapse
determines the population of objects at the corresponding scale. If this is
the case, the index $\alpha $ for the kernel ${K_{\mathbf{k,}\alpha \gamma }}
$ (see just below Eq.(\ref{smoluchowski})) should be about $-1.25-\left(
-1\right) =0.25$. The exponential factor $\exp \left( -\frac{L}{L_{\ast }}%
\right) $ seems to be a natural tail of the distribution which is gradually
evolving without runaway coalescence.

\item We could not complete the argument on the velocity-luminosity (or
mass) relation [38] which holds within a single species of object, from our
point of view. If we simply apply our argument for luminous objects, we
would have, from $\rho _{LM}=\rho _{LM0}r^{-1.8}$, the luminosity 
\begin{equation}
L=\left( \frac{L_{\ast }}{M_{\ast }}\right) \frac{4\pi }{3}\rho
_{LM0}\varepsilon ^{-1.2}v^{3.6}  \label{Lv-rel}
\end{equation}%
as a function of the velocity $v$ or the velocity dispersion $\sigma $\ at
that scale, provided appropriate reference ratio $L_{\ast }/M_{\ast }$ is
given. For DM, we have, from $\rho _{DM}=\rho _{DM0}r^{-4/3}$, the mass
expression 
\begin{equation}
M=\frac{4\pi }{3}\rho _{DM0}\varepsilon ^{-5/3}v^{5}.  \label{Mv-rel}
\end{equation}%
On the other hand, several tight relations are obtained for each type of
galaxies. For spiral galaxies, Tally-Fisher relation of the form Eq.(\ref%
{Lv-rel}) holds with the variation of index from 3 for B-band to 4 for K
band observations. For elliptical galaxies, a tight relation on the
fundamental plane is established, $r=$const.$\sigma ^{1.24}I^{-0.82}$ where $%
I=L/\left( 2\pi r^{2}\right) $ is the average surface brightness. Although
we can simply apply our argument to yield 
\begin{equation}
r^{1.07}=\text{const.}\sigma ^{1.24}I^{-0.82},
\end{equation}%
which is almost consistent with the above, we cannot obtain the relation in
the form of a plain among three independent parameters; all quantities are
functions of only $r$ in our case.
\end{enumerate}

We would like to elaborate our argument along the above issues and hope we
can report the results soon.

\bigskip

{\Large Acknowledgements}

The authors would like to thank Osamu Hashimoto, Hidenori Takahashi for
relevant discussions on the turbulence, Osamu Iguchi, Yasuhide Sota, and
Tohru Tashiro for discussion on the local virial relation, Nozomi Mori for
the cosmic magnetic fields and dynamo theory, Masaaki Morita, Hideaki Mouri
for turbulence and fate of cosmic vorticity, Kiyotaka Tanikawa for showing
us his paper before publication and providing us relevant references on $J-M$
relation, Takayuki Tatekawa for higher-order density perturbation theories.
All of their help has been crucial for the completion of this paper.

\end{document}